\documentclass[journal]{IEEEtran}
\usepackage{graphicx}
\usepackage{array}
\usepackage{mdwmath}
\usepackage{mdwtab}
\usepackage{eqparbox}
\usepackage{algorithmic}
\usepackage[cmex10]{amsmath}
\usepackage{setspace}
\usepackage{algorithm,caption}
\usepackage{empheq}
\usepackage{subcaption}
\usepackage{cases}
\usepackage{soul}
\usepackage{color}
\usepackage{cite}
\usepackage{amssymb}
\usepackage{multirow}
\usepackage{url}

\allowdisplaybreaks
\ifCLASSINFOpdf

\else

\fi
\begin{document}
%
\title{Towards Ultra-Reliable Low-Latency Communications: Typical Scenarios, Possible Solutions, and Open Issues}

\author{Daquan~Feng$^{\dag}$, Changyang She$^{\ddag}$, Kai Ying$^{\S}$, Lifeng~Lai$^{\dag}$, Zhanwei~Hou$^{\ddag}$, Tony~Q.~S.~Quek$^{\flat }$, Yonghui~Li$^{\ddag}$, and Branka~Vucetic$^{\ddag}$
\\
\IEEEauthorblockA{$^{\dag}$\, Guangdong Key Laboratory of Intelligent Information Processing, Shenzhen University, China\\
$^{\ddag}$\,School of Electrical and Information Engineering,
University of Sydney, Sydney, Australia\\
$^\S$\,Sharp Laboratories of America, Camas,
WA, USA\\
$^{\flat}$\, ISTD Pillar, Singapore University of Technology and Design, Singapore\\
}
\thanks{$^*$\,Accepted by IEEE Vehicular Technology Magazine, 2019.}
}
\maketitle

\begin{abstract}
\emph{Ultra-reliable low-latency communications} (URLLC) has been considered as one of the three new application scenarios in the \emph{5th Generation} (5G) \emph {New Radio} (NR), where the physical layer design aspects have been specified. With the 5G NR, we can guarantee the reliability and latency in radio access networks. However, for communication scenarios where the transmission involves both radio access and wide area core networks, the delay in radio access networks only contributes to part of the \emph{end-to-end} (E2E) delay. In this paper, we outline the delay components and packet loss probabilities in typical communication scenarios of URLLC, and formulate the constraints on E2E delay and overall packet loss probability. Then, we summarize possible solutions in the physical layer, the link layer, the network layer, and the cross-layer design, respectively. Finally, we discuss the open issues in prediction and communication co-design for URLLC in wide area large scale networks.
\end{abstract}


\IEEEpeerreviewmaketitle

\section{Introduction}
The Fifth Generation (5G) New Radio (NR) considers three new application scenarios, namely enhanced Mobile Broadband (eMBB), massive Machine-Type Communications (mMTC), and Ultra-reliable Low-latency Communications (URLLC) \cite{3GPP2017Scenarios}. URLLC is crucial for enabling mission-critical services, such as factory automation, automation vehicles, remote control and virtual/augmented reality (VR/AR). 

There are many open technical hurdles ahead in achieving URLLC, and thus it has attracted significant attention from both the academic and industrial communities. In the current Long Term Evolution (LTE) systems, the transmission time interval (TTI) is $1$~ms, which cannot satisfy the end-to-end (E2E) delay requirement of URLLC. To reduce the latency, short frame structure with short channel codes should be considered. With short codes, it is very difficult to achieve the ultra-high reliability requirement. Analyzing and optimizing the transmission delay and the decoding error probability in the short blocklength regime are also very challenging \cite{Yury2010Channel}.

Aside from transmission delay and decoding error probability, other delay components and delay bound violation probabilities in scheduling procedure and queueing systems also have significant impacts on the E2E performance. For example, in LTE systems, the control signaling for uplink (UL) scheduling leads to a high latency that is much longer than 1 ms in the control plane \cite{Condoluci2017Soft}. Thus, how to design grant-free access techniques for URLLC deserves further study. Besides, with the first-come-first-serve (FCFS) scheduling policy, the short packets of URLLC services may need to wait for the processing of long packets of eMBB services. Thus, FCFS policy may not be the optimal policy for short packets in URLLC services, and other policies should be considered to minimize the E2E delay.

The current techniques in the 5G NR \cite{3GPP2017Agree} mainly focus on achieving the target E2E performance in local area communications, where all the user equipment (UE) lies in one or few adjacent cells. 
For different communication scenarios, the network architectures are different. In factory automation, the communication area is limited in a smart factory, while for remote control, the controller and slave can be located on different continents. As a result, the latency in radio access network only contributes a small portion of the E2E delay, and other delay components such as core network delay over a long distance large scale network and processing delay in the computing systems may be the dominant components \cite{aijaz2017realizing}. Therefore, how to improve the E2E performance with different network architectures is still a challenging issue.

In this paper, we focus on how to guarantee the E2E delay and overall packet loss probability in different communication scenarios, including local area communications, mobile edge computing (MEC) systems, and the long distance large scale networks. The rest of this paper is organized as follows:
\begin{itemize}
	\item  We elaborate possible components of the E2E delay and overall packet loss probability in typical communication scenarios, and provide a general way to formulate the quality-of-service (QoS) constraints of URLLC.
	\item  We summarize possible solutions and techniques in physical layer, link layer, network layer, and cross-layer design aspects for URLLC, such as 5G NR physical layer technologies, different packet scheduling policies, and network slicing.
	\item We outline the basic idea in prediction and communication co-design for URLLC in long distance large scale networks and discuss some open issues.
\end{itemize}


\section{E2E Delay and Overall Packet Loss Probability}
\begin{figure}[htbp]
        \centering
        \begin{minipage}[t]{0.5\textwidth}
        \includegraphics[width=0.9\textwidth]{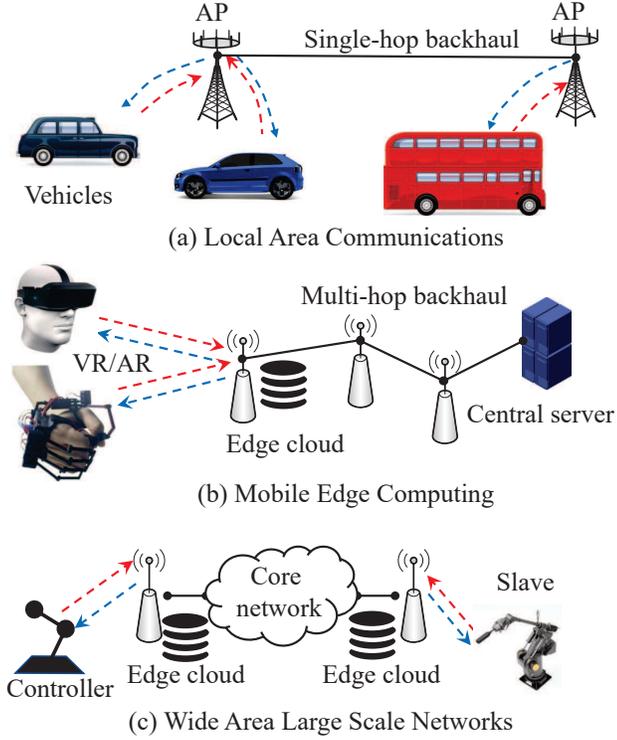}
        \end{minipage}
        \caption{Three typical communication scenarios for URLLC services.}
        \label{fig:components}
        \vspace{-0.0cm}
\end{figure}

The delay components and factors that lead to packet loss depend on network architectures. In this section, we first discuss them in three typical communication scenarios illustrated in Fig. \ref{fig:components}: local area communications, mobile edge computing, and wide area large scale communications. Then, we provide a general way to formulate QoS constraints of URLLC.

\subsection{Local Area Communications}
In local area communications, all the UEs are served by a few adjacent access points (APs) that are interconnected by single-hop fiber backhaul. One typical application requiring only local area communications is vehicle safety applications, where safety messages are shared among close-located vehicles. In this scenario, the E2E delay includes UL and downlink (DL) transmission delays, $D_{\rm t}$, queueing delay in the buffer of the APs, $D_{\rm q}$, and UL access delay, $D_{\rm a}$, while the propagation delay and backhaul delay are negligible since they are much smaller than $1$~ms. 

To achieve low-latency, the transmission delay should be short and thus it requires short blocklength channel codes. To achieve ultra-high reliability, the decoding error probability in the short blocklength regime, $\varepsilon_{\rm t}$, cannot be ignored \cite{Yury2010Channel}. Besides, a packet will become useless, if the queueing delay or access delay violates the corresponding delay bounds. Therefore, the queueing delay violation probability, $\varepsilon_{\rm q}$, and access delay violation probability, $\varepsilon_{\rm a}$, should be considered for URLLC services.

\subsection{Mobile Edge Computing}
In smart factories or VR/AR applications, UEs may not have sufficient processing capability. In this case, to reduce processing delay at the local computing system, UEs can offload tasks to MEC systems. In MEC systems, all the delay components and packet loss factors in the local area communications should be considered. Besides, the processing delay, $D_{\rm p}$, could be comparable to other delay components. Moreover, if packets are sent to central servers via multi-hop backhaul, then the backhaul delay, $D_{\rm b}$, may be dominant. Similar to queueing and access procedure, if the processing of a packet is not finished in time or the backhaul delay violates the required delay bound, the packet is lost. Thus, both the processing delay violation probability, $\varepsilon_{\rm p}$, and the backhaul delay violation probability, $\varepsilon_{\rm b}$, should also be taken into account in MEC systems.

\subsection{Wide Area Large Scale Networks}
Different from the traditional internet that supports real-time audio and video communications, some remote control applications aim to deliver real-time control and tactile feedback (e.g., industrial control, remote driving or tele-robotic surgery.). As stated in \cite{aijaz2017realizing}, the long-term ambition of Tactile Internet is to enable the sharing of skills globally. In wide area core networks, additional delays are incurred due to intermediate data center/cloud. In this case, the overall latency is dictated not just by the radio access networks, but also the backhauls, the wireless core networks, and processing in data center. For example, if the distance between the controller and the slave is $3000$~km, the propagation delay, $D_{\rm g}$, is around $10$~ms. To handle this issue, one promising solution is to deploy intelligent MEC to predict the mobilities of the controller and the slave, and transmit their control and feedback information in advance \cite{Mischa2018Towards}.

\subsection{Constraints on E2E Delay and Overall Packet Loss Probability}
Denote the requirement of the E2E delay and overall packet loss probability as $D_{\max}$ and $\varepsilon_{\max}$, respectively. Then, the delay and reliability can be satisfied under the following two constraints,
\begin{align}
D_{\rm t} + D_{\rm q} + D_{\rm a} + D_{\rm p} + D_{\rm b} + D_{\rm g} &\leq D_{\max},\label{eq:E2ED}\\
(1-\varepsilon_{\rm t})(1-\varepsilon_{\rm q})(1-\varepsilon_{\rm a})(1-\varepsilon_{\rm p})(1-\varepsilon_{\rm b}) &\leq  1-\varepsilon_{\max}.\label{eq:E2Eepsilon}
\end{align}
In the following, we discuss the recent advances in the physical layer, link layer, network architecture, as well as cross-layer design to ensure the requirements in \eqref{eq:E2ED} and \eqref{eq:E2Eepsilon}. The typical applications, possible solutions, and open issues are summarized in Table \ref{T:summary}.

\begin{table*}[htbp]
\small
\renewcommand{\arraystretch}{1.3}
\caption{Applications, possible solutions, and open issues in different communication scenarios}
\begin{center}\vspace{-0.2cm}\label{T:summary}
\begin{tabular}{|p{2.3cm}|p{3.7cm}|p{4.5cm}|p{5cm}|}
  \hline
  Communication Scenarios & Applications & Possible Solutions & Open Issues  \\\hline
  Local areas communications& Road safety applications and autonomous vehicles \cite{3GPP2017Scenarios}& 5G NR, grant-free access, and multi-connectivity & Analyzing overhead for channel estimation and correlation of shadowing \\\hline
  Edge computing systems & Virtual/augmented reality and factory automation \cite{Mischa2018Towards} & Improving scheduling scheme in communication and computing systems & Optimizing communication and computing systems and characterizing E2E delay and reliability\\\hline
  Wide area large scale networks & Health care, remote control, and smart grid \cite{aijaz2017realizing}& Prediction \& communication co-design & Designing accurate prediction method and jointly optimizing prediction \& communication systems   \\\hline
\end{tabular}
\end{center}
\vspace{-0.2cm}
\end{table*}

\section{PHYSICAL LAYER TECHNOLOGIES}
Physical layer design is among the most challenging and important issues for the three communication scenarios of URLLC applications. For URLLC in 5G NR, the target user plane latency is 0.5 ms each way for both UL and DL while the target reliability is $99.999$\% success probability for transmitting a packet of 32 bytes  within 1ms \cite{3GPP2017Scenarios}.


\subsection{Flexible Numerology and Frame Structure}
In the current 4G networks, the TTI is $1$ ms. Thus, the TTI should be shortened to meet the latency requirement of URLLC. From frame structure point of view, there are two ways to shorten the TTI. One is to increase the subcarrier spacing (SCS) so that the symbol duration can be decreased. In 5G NR, the SCS is flexible, which is given by $\triangle f=2^{\mu}\cdot 15$~kHz and $\mu= 0, 1, 2, 3, 4$. Thus, the TTI can be reduced by selecting larger SCS. For example, when the number of OFDM symbols in a slot is fixed as $14$, the slot duration with $30$~kHz SCS is $0.5$~ms, while the slot duration with $60$~kHz SCS is $0.25$~ms.

The other way to shorten the TTI is to reduce the number of OFDM symbols in a TTI. That is the motivation to introduce mini-slot in 5G NR. The number of OFDM symbols in a mini-slot can be $\{2,4,7\}$. As a result, the TTI can be further shortened. For example, the duration of a 2-symbol mini-slot with $30$~kHz SCS is $71.4$~us, which is much shorter than that the TTI in LTE.

\subsection{Self-Contained Slot Structure}

In the time division duplex (TDD) mode, when an AP receives a scheduling request from a UE, it has to wait until next available DL slot to send out a UL grant. However, if it is an UL-heavy configuration, there are fewer DL slots, then the waiting time can be very long. Similarly, in a DL-heavy configuration, a quick acknowledgment to a DL data reception may not be available.

Thus, in 5G NR, self-contained slot structure is introduced, where OFDM symbols in a slot can be classified as DL, flexible, or UL. In other words, both directions can be supported within a single slot. In this case, with the help of self-contained slot structure, the waiting time in TDD systems can be shortened effectively. For example, after receiving a DL data at the beginning of a slot, UE can feedback the corresponding acknowledge at the end of the same slot.

%
%
%

\subsection{CQI and MCS Table for URLLC}
To guarantee the reliability of data transmission, an appropriate modulation and coding scheme (MCS) according to the channel quality indication (CQI)  should be selected from a look-up table to meet the block error rate (BLER) target, i.e., the decoding error probability in Section II. The BLER target of eMBB is set as $10^{-1}$, which is the same as LTE. For URLLC, the target BLER is below $10^{-6}$ \cite{3GPP2017Scenarios}.

On the other hand, to achieve a successful data transmission, the control message must be reliable, no matter whether it is for resource assignment or feedback. Intuitively, there are two basic ways to enhance the control reliability. One is to enlarge the control resource and the other is to shorten the size of control information. Both ways help to encode the control message with a low coding rate so that the reliability is enhanced.


\subsection{Slot Aggregation and Repetition}
To improve reliability, we introduce slot aggregation (for grant-based transmission) and repetition (for grant-free transmission) in NR. The basic idea of slot aggregation and repetition is that an initial transmission of a packet can be followed by automatic repetitions of the same packet in consecutive slots.
The aggregation factor (or the number of repetitions) $K$ is configured by the higher layer. $K$=1 means there is no aggregation (or repetition) after the initial transmission. According to the current NR specification, the largest value of $K$ is $8$, which is large enough to guarantee the ultra-reliability of the data transmission. On the other hand, from the latency perspective, the retransmission timeline is reduced by using repetitions. Retransmission is always grant based, which is time-consuming. However, repetitions are automatically transmitted in the consecutive slots without waiting for any grant or retransmission feedback, and thus help to reduce latency.

\section{Link Layer Design}
In the section, we focus on link layer design and consider the scenarios with random packet arrival processes. To reduce latency, grant free access for UL transmission and DL scheduling policies in communication and processing systems are summarized. In addition, to improve reliability, D2D communications, relay systems, cellular links, and multi-connectivity are discussed.

\subsection{Grant Free Access for UL Transmission}
With the current LTE protocol, when a UE has a packet to transmit, it first uploads a scheduling request to the AP. Then, the AP sends a transmission grant to the UE. Finally, the UE can upload its packet. Such an UL scheduling procedure lead to long access delay. To reduce access delay, grant-free access has been proposed as a promising solution.

Reserving dedicated bandwidth for each UE is a natural way to avoid access delay. However, such a method is only suitable for UEs with high packet arrival rate. Reserving bandwidth for each UE with low packet arrival rate leads to very low bandwidth usage efficiency. To address this issue, a contention-based access procedure has been studied in \cite{singh2017contention}, i.e., the slotted ALOHA access scheme. With the contention-based access, the total bandwidth is divided into multiple channels. In each slot, UEs that need to send packets, choose one of the channels randomly for data transmission. If more than one UEs choose the same channel, then the transmissions will fail.

\begin{figure}[htbp]
        \centering
        \begin{minipage}[t]{0.5\textwidth}
        \includegraphics[width=0.9\textwidth]{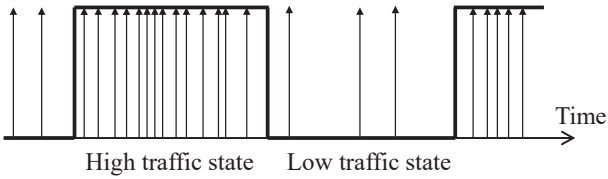}
        \end{minipage}
        \caption{Bursty packet arrival process.}
        \label{fig:bursty}
        \vspace{-0.0cm}
\end{figure}
According to the experiment in \cite{Condoluci2017Soft}, the arrival processes in some applications are very bursty. As illustrated in Fig. \ref{fig:bursty}, the arrival rate of a bursty arrival process switches between a high traffic state and a low traffic state. To avoid high collision probability and to save bandwidth, the authors in \cite{hou2018Burstiness} first classify the arrival processes into the high and low traffic states. Then, dedicated bandwidth is reserved for UEs in the high traffic state and the slotted ALOHA access scheme is applied for UEs in the low traffic state. To guarantee the reliability requirement, the classification errors should be considered \cite{hou2018Burstiness}.

\subsection{DL Scheduling Policies in Communication and Computing Systems}
\begin{figure}[htbp]
        \centering
        \begin{minipage}[t]{0.5\textwidth}
        \includegraphics[width=0.9\textwidth]{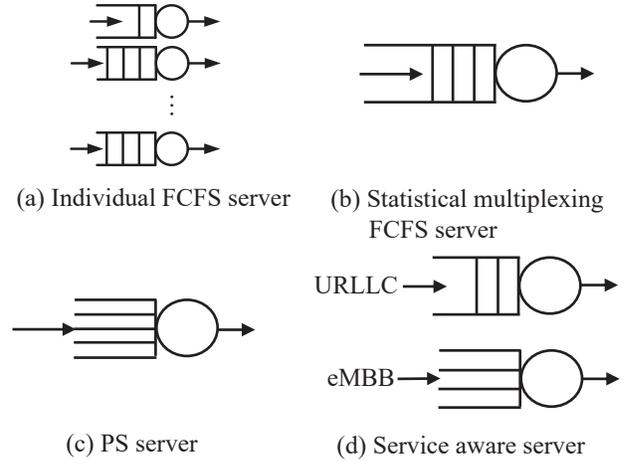}
        \end{minipage}
        \caption{Typical scheduling policies.}
        \label{fig:scheduling}
        \vspace{-0.0cm}
\end{figure}
In traditional communication systems, the packets to different destinations are waiting in different queues at the buffer of the AP, i.e., the individual FCFS server in Fig. \ref{fig:scheduling}(a). Such a policy can guarantee the QoS of each user. However, the resource utilization efficiency is low because the resources allocated to different users cannot be shared among each other even some users' queues are empty. To improve resource utilization efficiency, the statistical multiplexing server in Fig. \ref{fig:scheduling}(b) can be used, where packets of different users stay in one queue. As proved in \cite{ResourceShe}, if the arrival process of each user follows Poisson process and the packet size are identical (e.g., each short packet in URLLC contains $20$~bytes \cite{3GPP2017Scenarios}), then the statistical multiplexing server can guarantee the QoS of all the packets with less total bandwidth compared with that adopting individual server. 

In practical systems, the transmission/processing time of different kinds of packets can be very diverse. Since the packet size of eMBB services is much larger than URLLC services, the transmission/processing time for the long packets in eMBB services is much longer than the short packets in URLLC services. As a result, if we use the FCFS servers, the short packets that arrive at the server following a long packet, need to wait for a long time. To avoid this situation, one solution is to use the Processor-Sharing (PS) server as illustrated in Fig. \ref{fig:scheduling}(c). In the PS server, the total service ability of the server is equally allocated to all the packets in the buffer \cite{Mor2013Queue}. In this way, the short packets do need to not wait for the processing of long packets. Furthermore, if the server is aware of the packets in different services, then it is possible to design different scheduling policies for different kinds of services (e.g., the service aware server in Fig. \ref{fig:scheduling}(d)). 

\subsection{D2D, Relay, and Cellular Links for URLLC}
In some applications of URLLC like factory automation and vehicle networks, each device transmits short packets to nearby devices. For these short distance communication scenarios, D2D communications may outperform cellular links. However, when using D2D communication in the URLLC scenario, interference should be avoided. One possible solution is using APs to manage radio resources, and sending data packets via D2D links.

Considering that D2D communications have limited communication range, the relay systems are applied in \cite{Yulin2018Relaying}. The results in \cite{Yulin2018Relaying} show that the relay systems can achieve higher throughput or lower queueing delay in comparison to direct transmission in both noise-limited and interference-limited scenarios.

To further improve reliability, one important technique is multi-connectivity \cite{AvailableRangeShe}. The basic idea is transmitting one packet overall multiple parallel links, such as D2D links, relay, and cellular links. The results in \cite{AvailableRangeShe} show that the achieved reliability decreases with the cross-correlation of shadowing among parallel links. However, how to analyze the impact of cross-correlation of shadowing on reliability remains an open problem.

\section{Novel Network Layer Design}
The existing cellular network architecture is mainly designed to meet the requirements for conventional mobile broadband services. However, it can not support the diversified 5G services. Thus, the novel mobile computing frameworks and network architecture techniques, such as network slicing, software-defined networking (SDN), network function visualization (NFV), and self-organizing networks (SON) are attracting considerable attention. In this section, we introduce the recent advances of these techniques for URLLC services.

\subsection{Network Slicing with SDN/NFV}
Network slicing is a fundamental technology for the future networks to provide diversified services simultaneously over the same physical infrastructure \cite{ksentini2017toward}. It allows the network to build  multiple logical sub-networks with reserved resources for different application scenario, e.g., eMBB, mMTC, and URLLC. In this context, the URLLC service can avoid the interruptions by the other services that share the same resource and thus help to enhance QoS and user experience.  However, due to the traffic fluctuations and channel variability, strict isolation and sharing between multiple slices faces a great challenge and it may lead to low resource utilization efficiency.

SDN and NFV are two pillars to support multi-virtual networking slicing in 5G. Particularly, SDN separates the control plane from the forwarding plane and offers the centralized network flow management to simplify the scheduling and resource allocation. On the other hand, NFV decouples network functions from dedicated hardware to provide programmability and flexibility over the entire network. With the integration of  SDN and NFV,  network operators can provide efficient, scalable, and flexible network slice service configuration on demand. Therefore, network slicing with SDN and NFV can significantly improve overall performance of networks, including delay and reliability in radio access networks, backhauls, and core networks \cite{URLLC2018Survey}. In \cite{Slicing2}, it is shown that the SDN-based network architecture can achieve up to $75$\% performance improvement in E2E latency. In \cite{Slicing3}, the authors propose two-level MAC scheduling framework for a slicing-enabled 5G network. It is shown that with dynamic slice management, the stringent requirements for URLLC can be guaranteed.


\subsection{Reducing Latency with MEC}
MEC is also considered as a promising solution to reduce latency for processing tasks of URLLC services. In \cite{mec1}, MEC is considered to integrate with SDN and NFV to deal with the service disruption incurred by user mobility. It is demonstrated that distributed and virtualized network provisioning can effectively reduce latency and improve resiliency. In \cite{MECTaskOffloading}, the trade-off between power and delay in MEC is studied, where computation and transmit power is minimized by optimizing task-offloading and resource allocation. Since the FCFS scheduling policy is considered in this paper, the latency of short packets is not optimized. How to optimize task-offloading and scheduling policy subject to the QoS requirements of URLLC, eMBB, and mMTC deserves further study.

\subsection{SON}
Networking slicing, SDN and NFV can achieve better network scalability and flexibility. However, they also result in much more complicated management and configuration of the network, which may depress the QoS and user experience. Thus, it is critical to adopt the SON management mechanisms to provide intelligence, automatic, and distributed management and optimization \cite{View20175GPPP}. By taking advantage of the rapid progress of big data processing and machine learning technologies, the EU 5G-PPP project propose a catalog-driven network management system to enable smart deployment of service. Nevertheless, how to guarantee the QoS requirement of URLLC in SON deserves further study.

\section{Improving E2E Performance with Cross-layer Design}

Considering that each layer of the protocol stack has an inherent interdependence on other layers, cross-layer resource management has the potential to improve the E2E delay and overall reliability. For example, the transmission delay, queueing delay, and routing delay depend on physical layer, link layer, and network layer, respectively. By optimizing the delay components subject to the E2E delay constraint in \eqref{eq:E2ED} we can achieve better resource utilization efficiency. In this section, we will illustrate how to save bandwidth or transmit power with cross-layer design.

\subsection{A Different Conclusion Obtained from Cross-layer Design}
\begin{figure}[htbp]
        \centering
        \begin{minipage}[t]{0.45\textwidth}
        \includegraphics[width=1\textwidth]{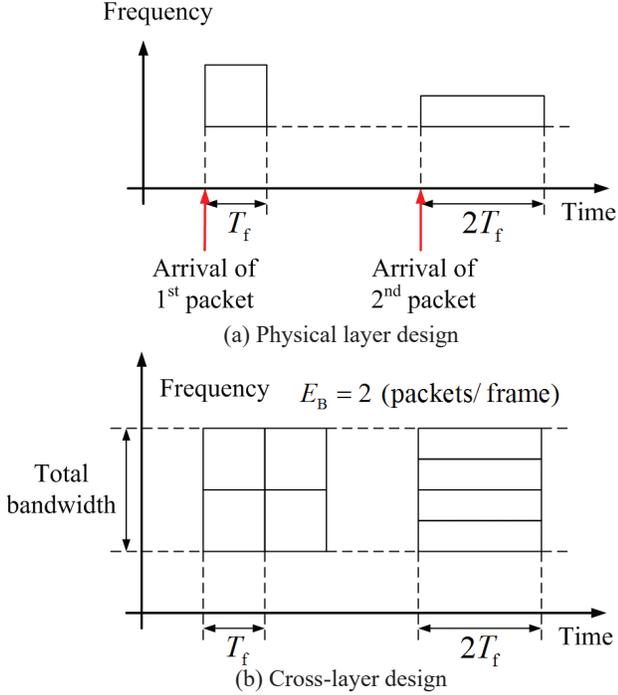}
        \end{minipage}
        \caption{Illustration on why cross-layer design is necessary.}
        \label{fig:crosslayer}
        \vspace{-0.0cm}
\end{figure}

In physical layer design, it is well-known that there are tradeoffs among physical layer resources, e.g., transmission time, bandwidth, and transmit power. As illustrated in Fig. \ref{fig:crosslayer} (a), if we double the transmission duration, then only half bandwidth is required if the rate of the channel code remains constant. Besides, the transmit power to achieve the same SNR is also halved.

However, the conclusion is different from a cross-layer perspective. We consider a FCFS queueing system, and assume that to guarantee a queueing delay bound, $D_{\rm q}$, and queueing delay violation probability, $\varepsilon_{\rm q}$, the required service rate, referred to as effective bandwidth \cite{ResourceShe}, is $E_{\rm B} = 2$ (packets/frame). If the transmission duration of each packet is $1$~frame, then $2$ packets are transmitted simultaneously. As shown in Fig. \ref{fig:crosslayer} (b), if the transmission duration of each packet is $2$~frames, then to achieve the required service rate, i.e., $2$~(packets/frame), the packets that are transmitted simultaneously is $4$. As a result, although the bandwidth for each packet is halved, the total bandwidth and total transmit power remain unchanged. Therefore, increasing the transmission duration does not help reduce bandwidth or transmit power, but leads to extra transmission delay. Consequently, the optimal transmission duration that minimizes the required bandwidth subject to the constraints on transmission and queueing delays and overall packet loss probability is $1$ frame \cite{ResourceShe}.

\subsection{Useful Insights in Cross-layer Design}
\begin{figure}[htbp]
        \centering
        \begin{minipage}[t]{0.5\textwidth}
        \includegraphics[width=1\textwidth]{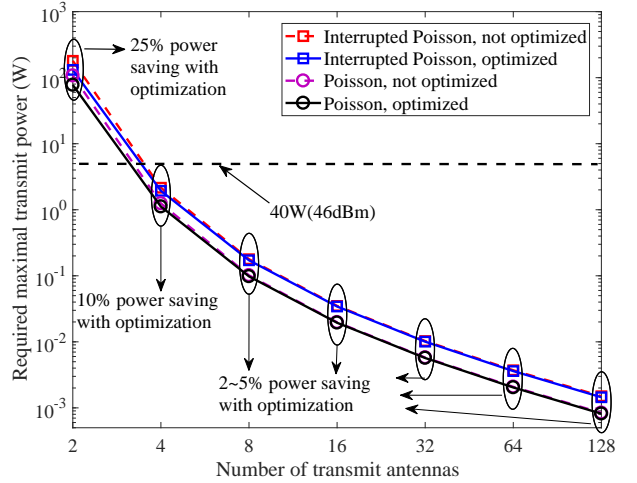}
        \end{minipage}
        \caption{Performance gain by optimizing packet loss probabilities \cite{Cross-Layershe}.}
        \label{fig:optE}
        \vspace{-0.0cm}
\end{figure}

\begin{figure}[htbp]
        \centering
        \begin{minipage}[t]{0.5\textwidth}
        \includegraphics[width=1\textwidth]{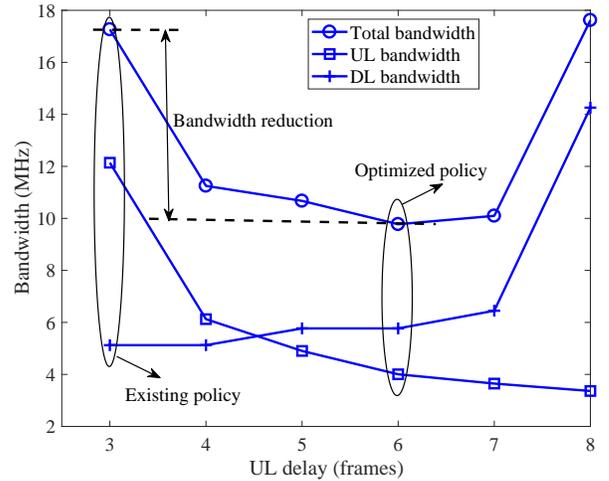}
        \end{minipage}
        \caption{Performance gain by optimizing delay components \cite{ResourceShe}.}
        \label{fig:optD}
        \vspace{-0.0cm}
\end{figure}

In radio access network, the most challenging issue in cross-layer design is how to obtain the optimal solution subject to the requirements on the transmission and queueing delays and different packet loss probabilities. In \cite{Cross-Layershe}, the required transmit power is minimized by jointly optimizing the probabilities of decoding error, queueing delay violation, and packet dropping over deep fading wireless channels subject to the overall packet loss probability requirement. The results in Fig. \ref{fig:optE} indicate that only $2\sim5$\% power gain can be obtained by optimizing the packet loss probabilities when the number of antennas at the AP is larger than $8$. A near optimal solution is setting all the packet loss components in \eqref{eq:E2Eepsilon} as equal. Furthermore, the uplink and downlink transmission delays and queueing delay are optimized subject to the E2E delay requirement in \cite{ResourceShe}. The results in Fig. \ref{fig:optD} show that by optimizing these three delay components, around half bandwidth can be saved. This is because the required resources are very sensitive to the delay components, but are less sensitive to the packet loss probabilities.

\section{Towards Wide Area Large Scale Networks: Prediction and Communication Co-design}
\begin{figure}[htbp]
        \centering
        \begin{minipage}[t]{0.5\textwidth}
        \includegraphics[width=0.9\textwidth]{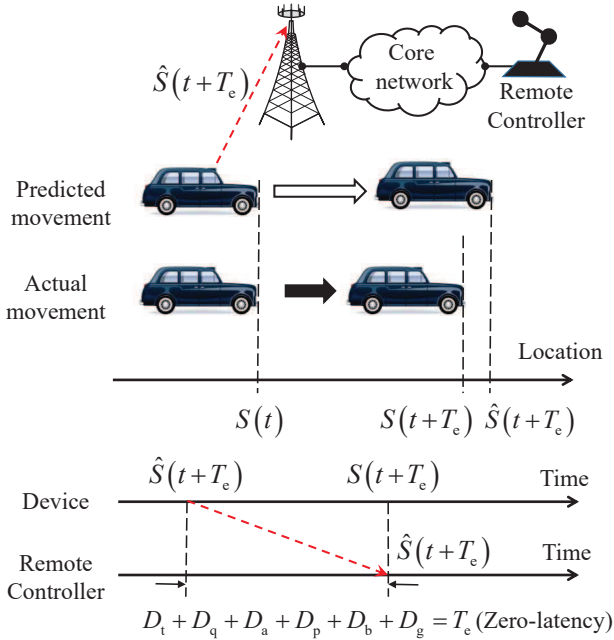}
        \end{minipage}
        \caption{Illustration of Prediction and Communication.}
        \label{fig:prediction}
        \vspace{-0.0cm}
\end{figure}

The propagation delay, $D_{\rm g}$, will be higher than $1$~ms as long as the communication distance is longer than $300$~km. Thus, it is impossible to achieve $1$~ms E2E delay only with physical layer technologies. Inspired by existing studies on mobility prediction \cite{dandapat2013sprinkler,zhang2015edgebuffer,ozfatura2018mobility}, we propose prediction and communication co-design method to handle this issue. The basic idea is predicting the movement of the device and send the predicted information in advance. Assuming that the system can predict the mobilities of the controller and the slave, $T_{\rm e}$ seconds, in advance, the delay in the core network experienced by the controller and slave can be reduced. In this case, prediction errors will lead to packet loss, and we denote $\varepsilon_{\rm c}$ as the packet loss probability due to prediction errors. The UL transmission with prediction is illustrated in Fig. \ref{fig:prediction}. At time $t$, the device predicts it's future location $\hat{S}(t+T_{\rm e})$ and sends the predicted information to the remote controller. With communication delay, the E2E delay experienced by the remote controller is the same as the right-hand side of \eqref{eq:E2ED}. If the sum of the delay components in the communication system equals to the prediction time, then it is possible to achieve zero-latency.

However, there are three open issues: 1) Intuitively, there is a tradeoff between the prediction time, $T_{\rm e}$, and the prediction error probability, $\varepsilon_{\rm c} = \Pr\{|S(t+T_{\rm e}) - \hat{S}(t+T_{\rm e})|\}$. How to design an accurate prediction algorithm that achieves low prediction error probability with long prediction time deserves further study. Possible solutions include model-based methods with Markov Chain or first-order autoregressive model, and data-driven methods like linear regression and neural networks. 2) The latency in the communication system is a random variable depending on wireless channel fading, queueing, routing, and network congestion. How to satisfy the constraint on the probability that the experienced delay violating the delay bound is an open problem. 3) How to optimize the prediction time to minimize the overall packet loss probability for a given prediction algorithm and a communication system remains unclear. To this end, a prediction and communication co-design is necessary.

\section{Conclusion}
In this paper, we elaborated the delay components and packet loss probabilities in three typical communication scenarios for URLLC. Then, We summarize possible solutions and techniques in the physical layer, the link layer, and the network architecture design aspects for URLLC. The solutions from each of these three layers are important for enabling URLLC. However, without cross-layer optimization, the separated optimization in the three aspects cannot obtain the global optimal solution, and may lead to incorrect conclusions. Motivated by this fact, we presented some optimization results in cross-layer resource management. Finally, we outlined the basic idea in prediction and communication co-design for wide area large scale networks and discussed some open issues.
\ifCLASSOPTIONcaptionsoff
  \newpage
\fi

\section*{Acknowledgment}
This work was supported in part by the National Natural Science Foundation of China (NSFC) under Grant 61701317, Young Elite Scientists Sponsorship Program by CAST under Grant 2018QNRC001, Guangdong Natural Science Foundation under Grant 2017A030310371, Shenzhen Basic Research Program under JCYJ20170302150006125, The Start-up Fund of Shenzhen University under 2017076, Tencent ¡°Rhinoceros Birds¡± - Scientific Research Foundation for Young Teachers of Shenzhen University, The Start-up Fund of Peacock Project, the SUTD-ZJU Research Collaboration under Grant SUTD-ZJU/RES/01/2016 and SUTD-ZJU/RES/05/2016, ARC under Grant DP150104019 and DP190101988, and the University of Sydney.

\bibliographystyle{IEEEtran}
\bibliography{URLLC}

\end{document}